\documentclass{jpsj-suppl}
\usepackage{times}
\usepackage{color}

\usepackage{amsmath,amsthm,amssymb}
\usepackage{ascmac}
\usepackage{cite}

\recdate{September 20, 2013}

\title{
Cluster mean-field approach with density matrix renormalization group:
Application to the hard-core bosonic Hubbard model on a triangular lattice
}

\author{Ryota Suzuki\thanks{E-mail address: suzuki@stat.phys.titech.ac.jp}
 and Akihisa Koga}

\inst{Department of Physics, Tokyo Institute of Technology, Meguro, 
Tokyo 152-8551, Japan 
}
\abst{
We introduce a new numerical method for the solution of 
self-consistent equations in the cluster mean-field theory. 
The method uses the density matrix renormalization group method to solve
the associated cluster problem. 
We obtain an accurate critical value of the supersolid-superfluid transitions 
in the hard-core bosonic Hubbard model on a triangular lattice,
which is comparable with the recent quantum Monte Carlo results.
This algorithm is applicable to more general classes of models 
with a larger number of degrees of freedom.
}

\kword{cluster mean-field approach, density matrix renormalization group}

\begin{document}
\maketitle

\section{Introduction}
Ultracold bosonic gases have attracted current interest since
the successful observation of the Bose-Einstein condensation 
in $^{87}\rm Rb$ atoms~\cite{Rb}.
One of the interesting systems is a bosonic gas
in the optical lattice, where local particle correlations suppress
an itinerancy of atoms and yield the competition between the superfluid 
and solid states.
In fact, the phase transitions between them 
have been observed in the bosonic systems 
on the cubic~\cite{Greiner} 
and triangular lattices~\cite{Becker}.
On the other hand, it has theoretically been suggested 
that a coexistence between 
the superfluid and solid states, a so-called supersolid state, 
is realizable in some models~\cite{Matsuda,Mullin,Liu}.
In the hard-core bosonic model, it has been clarified that
the lattice geometry as well as intersite correlations play an important role
to stabilize the supersolid state~\cite{Boninsegni,Wessel2005,TSuzuki}.

In the hard-core bosonic system on the triangular lattice, 
the existence of the supersolid state has been clarified in terms of 
the quantum Monte Carlo (QMC) method~\cite{Boninsegni,Wessel2005}.
However, around the half filling, some quantum states compete with each other
and the nature of the phase transitions was not so clear.
Recently, it has been clarified that on the symmetric case,
a quantum phase transition is of second order, 
while it is of first order away from
half filling by means of various methods~\cite{Yamamoto2012,Bonnes2011,Zhang}.
Among them, the cluster mean-field (CMF) theory~\cite{Oguchi}
is one of the simple and efficient methods 
to study the nature of the phase transitions.
However, it is not so clear how the phase boundary depends on the cluster size
treated in the CMF method, which may be crucial to determine 
the second-order critical point.

In this paper, we introduce the density matrix renormalization group 
(DMRG) technique~\cite{White,DMRG,Nishino,Shibata} 
as a cluster solver.
We then deal with different clusters systematically in the CMF+DMRG method
to discuss the quantum phase transitions in the hard-core bosonic 
Hubbard model on the triangular lattice quantitatively.

The paper is organized as follows. In \S2, we introduce
the model Hamiltonian for the bosonic system on
the triangular lattice and summarize the CMF method with the DMRG technique. 
In \S3, we discuss the quantum phase transition between the supersolid
and superfluid states, and find that the obtained critical point is comparable
with the recent results obtained by the QMC method~\cite{Bonnes2011}.
A summary is given in the final section.

\section{Model and Method}
We consider zero-temperature properties in interacting bosons 
on the triangular lattice. 
Here, we assume sufficiently large onsite interactions. 
In the case, the system should be described by
the following hard-core bosonic Hubbard model as,
\begin{align}
  H=-t \sum_{\langle i,j \rangle} (\hat a_i^\dagger \hat a_j + \mathrm{h.c.})
  + V \sum_{\langle i,j \rangle} \hat n_i \hat n_j
  - \mu \sum_i \hat n_i,
  \label{hardcore-bose-hubbard}
\end{align}
where $\langle i,j\rangle$ denotes the summation over nearest neighbor sites,
$\hat a_i^\dagger (\hat a_i)$ is the creation (annihilation) operator 
at site $i$ and $\hat n_i (= \hat a_i^\dagger \hat a_i)$ is the number 
operator. $t$ is the hopping integral, $V$ is the intersite repulsion,
and $\mu$ is the chemical potential.

\begin{figure}[htb]
\centering
\includegraphics[width=10cm]{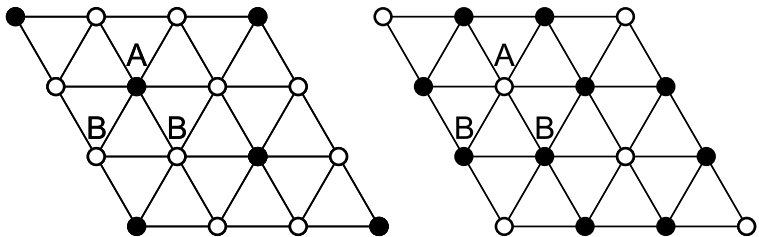}
\caption{Left (right) panel shows a density distribution 
of bosons in the solid state with $\rho=1/3 (\rho=2/3)$.
The lattice sites are devided into 
A (triangular) and B (honeycomb) sublattices.
}
\label{fig1}
\end{figure}

It is known that three kinds of ground states appears in the system
depending on the ratio $t/V$ and the filling 
$\rho(=\sum_i\langle \hat n_i\rangle/N)$,
where $N$ is the total number of sites.
When the interaction strength is small enough, 
the superfluid state with the order parameter,
$\Psi (= \sum_i \left< \hat a_i \right> / N)$, is realized.
On the other hand, in the strong coupling region,
the solid states with $\rho=1/3$ and $2/3$ are realized, where
the spatial distribution of bosons is schematically 
shown in Fig. \ref{fig1}.
These solid phases are characterized by the structure factor 
$S_\mathbf{Q} [= \sum_i \langle \hat n_i \rangle 
\exp(-i \mathbf{Q}\cdot\mathbf{r}) / N]$, where
$\mathbf{Q}=(4 \pi /3, 0)$.
Between these superfluid and solid states, 
two kinds of supersolid states appear with distinct fillings,  
where both order parameters ($\Psi$ and $S_\mathbf{Q}$) are 
finite~\cite{Wessel2005}.

To discuss the quantum phase transitions in the hard-core bosonic Hubbard
model on the triangular lattice quantitatively, we make use of the CMF method.
In the CMF method, the original lattice model is mapped 
to an effective cluster model, 
where particle correlations in the cluster
can be taken into account properly. 
The expectation values of the inter-cluster Hamiltonian are 
obtained via a self-consistency condition imposed on the effective cluster problem.
This method has an advantage in discussing quantum phase transitions correctly
since not only stable and metastable states but also unstable states
can be treated.
Therefore, the CMF method has successfully been applied to 
the quantum spin systems~\cite{Oguchi,Yamamoto2009} and 
bosonic systems~\cite{Hassan,Yamamoto2012}.
As for the hard-core bosonic Hubbard model on the triangular lattice,
the reasonable phase diagram has been obtained
by means of the CMF method with 
the exact diagonalization (ED)~\cite{Hassan,Yamamoto2012}.
However, around the second-order critical point,
the correlation length should diverge and 
the CMF+ED method with small clusters may not 
describe the critical phenomena.

\begin{figure}[htb]
\centering
\includegraphics[width=12cm]{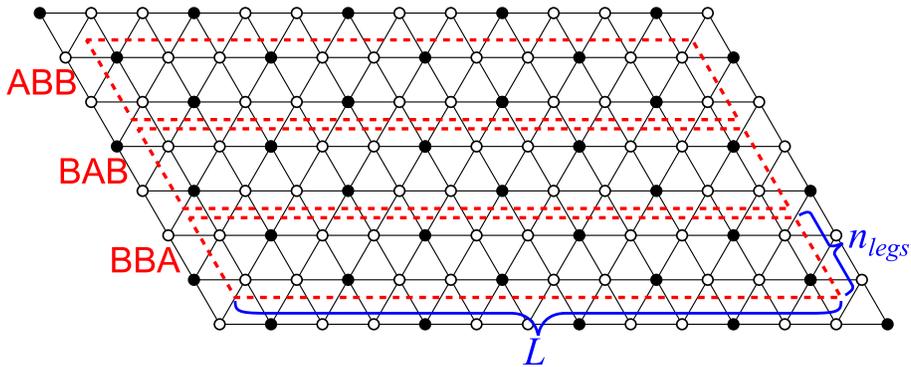}
\caption{(Color online) 
The effective ladder model with $n_{legs}\times L$ sites
for the CMF theory. 
Two-leg ladders with different density distributions are 
shown in the regions bounded by the dashed lines.
}
\label{fig2}
\end{figure}

To deal with larger clusters, 
we make use of the DMRG technique as an effective cluster solver.
It is known that this method is powerful 
for the one-dimensional systems~\cite{White,DMRG,Nishino,Shibata}.
Furthermore, by combining the DMRG method with a mean-field theory,
the phase transitions in the higher dimensions has been discussed 
in the Heisenberg models~\cite{Kawaguchi} and fermionic
Hubbard models~\cite{Maruyama,Garcia}. 
Here, using the DMRG technique, 
we solve the effective cluster model with a ladder structure 
$(n_{legs}\times L)$, where $n_{legs}$ is the number of legs and $L$ is 
the length of the ladder, as shown in Fig. \ref{fig2}.
The effective Hamiltonian is explicitly given as,
\begin{align}
H&=H_{intra}+H_{inter},\label{H2}\\
H_{intra}&=-t\sum_{( i,j )}
\left( \hat a_i^\dagger \hat a_j + \mathrm{h.c.}\right)
+V\sum_{( i,j )} \hat n_i \hat n_j ,\\
H_{inter}&=-t\sum_{( i,k )'}\left( \hat a_i^\dagger 
\langle \hat a_k \rangle + \mathrm{h.c.}\right)
+V\sum_{(i,k)'} \hat n_i \langle \hat n_k \rangle,
\end{align}
where the symbol $(i,j)$ [$(i,k )'$] denote the summation over 
nearest neighbor sites in the ladder (between ladders).
$\langle \hat n_k \rangle$ and $\langle \hat a_k \rangle$ are 
the expectation values of the number and annihilation operators 
at site $k$ in the nearest neighbor cluster.
In the paper, we introduce $[4 (L+n_{legs}-1)]$ mean-fields 
$\{\langle \hat n_k \rangle, \langle \hat a_k \rangle\}$.
By solving the cluster model and calculating the expectation values
by means of the DMRG method, we newly obtain mean-fields. 
In the CMF method, 
this iteration process is performed until these mean-fields are converged.
Note that in the DMRG calculations,
the quantum states $M$ kept in each step are limited
since the particle number does not conserve 
in the effective cluster model [eq. (\ref{H2})].
However, the large number of the quantum states $M$ is not needed
in the framework of the CMF method.
In fact, we did not find a visible difference of the results 
with $M=32$ and $M=64$.
Therefore, we fix the number of quantum states as $M=32$ 
in our CMF+DMRG calculations.


\section{Results}

We discuss the quantum phase transition between the supersolid and superfluid 
states on the symmetric line ($\mu/V=3$). 
Here, we focus on the order parameter characteristic of this phase transition
$\Delta (= \rho_{+}-\rho_{-})$~\cite{Yamamoto2012}, where
$\rho_\pm$ is the filling for the system with $\mu/V=3\pm\delta$, where
$\delta$ is infinitesimal.
In the superfluid state, the system is half-filled and $\Delta=0$. 
On the other hand, in the strong coupling region, 
two degenerate supersolid states
are realized with distinct fillings and thereby
$\Delta$ is finite.
In the following, we calculate this quantity
to discuss the quantum phase transition
between the supersolid and superfluid states.

\begin{figure}[htb]
\centering
\includegraphics[width=15cm]{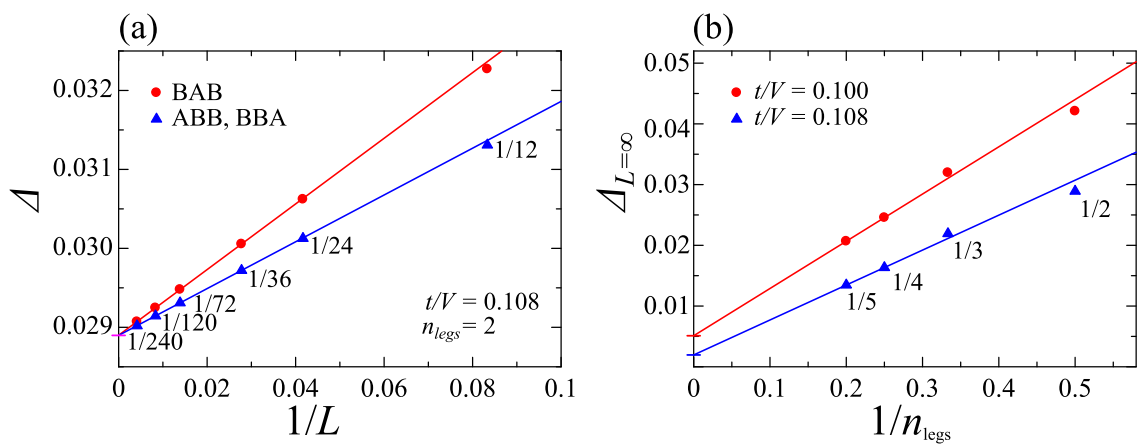}
\caption{(Color online) The system size dependence of the order parameter 
in the model with $\mu/V=3$. 
(a) the quantity $\Delta$ as a function of the inverse of the length $L$
in the two-leg ladders with $t/V=0.108$. 
Solid circles (triangles) represent the results
for the effective ladders with the BAB (ABB and BBA) structure 
(see Fig. \ref{fig2}).
(b) the quantity $\Delta$ as a function of the inverse of  
the number of legs $n_{legs}$ 
when $t/V=0.100$ (circles) and $0.108$ (triangles).
}
\label{fig3}
\end{figure}

In the CMF method, the scaling analysis for the data 
obtained from finite clusters is important to discuss critical phenomena.
First, we consider the length dependence of the data.
By solving the self-consistency equations of the CMF theory
for two-leg ladders with $L=12, 24, 36, 72, 120,$ and $240$, 
we obtain the results for $t/V=0.108$, as shown in Fig. \ref{fig3} (a).
It is found that the order parameter is well scaled 
by the inverse of the length $L$. 
We also confirm that the quantity for the ladder with $L\rightarrow\infty$  
does not depend on the choice of the clusters, as shown in Fig. \ref{fig2}.
Therefore, we can say that the order parameter for the ladder 
with the infinite length is obtained.
Next, we examine the scaling behavior for the number of legs of ladder, 
as shown in Fig. \ref{fig3} (b).
Here, we deduce the quantity in the thermodynamic limit 
$(n_{legs}\rightarrow \infty, L\rightarrow\infty)$
by assuming the scaling law $\Delta(n_{legs}) \sim \Delta + a/n_{legs}$.

\begin{figure}[htb]
\centering
\includegraphics[width=8cm]{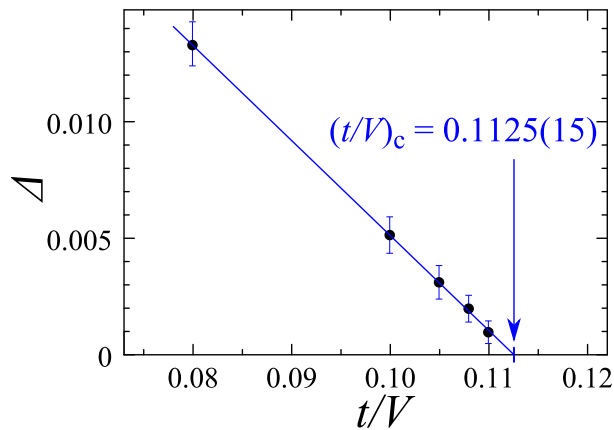}
\caption{(Color online) 
The order parameter $\Delta$ 
as a function of $t/V$ in the hard-core bosonic Hubbard system 
on the triangular lattice with $\mu/V=3$.}
\label{fig4}
\end{figure}

By performing similar calculations, we obtain the order parameter $\Delta$
in the thermodynamic limit, as shown in Fig. \ref{fig4}.
When $t/V$ is small, the order parameter is finite, implying that
the system is on the phase boundary between two supersolid states.
The increase in the hopping integrals between sites 
decreases the order parameter $\Delta$.
Finally, it vanishes and the second-order transition occurs
to the superfluid state.
The critical point is obtained as $(t/V)_c=0.1125(15)$,
which is in good agreement with the recent results obtained by 
the QMC method~\cite{Bonnes2011}.
On the other hand, this value is slightly larger than 
$(t/V)_c=0.108$ obtained by the CMF + ED method with 
small clusters~\cite{Yamamoto2012}.
This implies that the CMF+DMRG method with large clusters 
is more appropriate to discuss quantum phase transitions in
the hard-core bosonic system.

This algorithm is applicable to more general classes of models 
with a larger number of degrees of freedom.
One of the examples is a bosonic system on layered triangular lattices, 
which may be realized experimentally.
It is an interesting problem how the interlayer coupling affects the stability
of the supersolid states in the bosonic system, 
which is now under consideration~\cite{Suzuki}.

\section{Summary}
We have studied quantum phase transitions 
in the hard-core bosonic Hubbard model on the triangular lattice, 
combining the cluster mean-field theory
with the density matrix renormalization group.
Solving the effective Hubbard ladder model 
with two, three, four, and five legs systematically, 
we have extrapolated the order parameter for the supersolid-superfluid 
transition in the thermodynamic limit.
We have obtained the critical point at half filling $(t/V)_c=0.1125(15)$,
which is comparable with the recent results
obtained by the QMC method~\cite{Bonnes2011}.

\section*{Acknowledgments}
The authors would like to thank I. Danshita and D. Yamamoto 
for valuable discussions.
This work was partly supported by Japan Society for the Promotion of Science 
Grants-in-Aid for Scientific Research Grant Number 25800193
and the Global COE Program ``Nanoscience and Quantum Physics" from 
the Ministry of Education, Culture, Sports, Science and Technology (MEXT) 
of Japan.

\end{document}